\documentclass[sigconf, nonacm]{acmart}
\AtBeginDocument{%
  }

\copyrightyear{2025}
\acmYear{2025}
\setcopyright{rightsretained}
\acmConference[RecSys '25]{19th ACM Conference on Recommender Systems, Workshops}{September 22--26, 2025}{Prague, Czech Republic}
\acmBooktitle{19th ACM Conference on Recommender Systems (RecSys '25) Workshops, September 22-26, 2025, Prague, Czech Republic}

\usepackage{multirow}
\usepackage{algorithm}
\usepackage[noend]{algpseudocode}
\usepackage{amsmath}

\begin{document}

\title{ACT: Automated Constraint Targeting for Multi-Objective Recommender Systems}


\author{Daryl Chang, Yi Wu, Jennifer She, Li Wei, Lukasz Heldt}
\email{{dlchang, wuyish, jenshe, liwei, heldt}@google.com}
\affiliation{
  \institution{Google Inc}
  \city{Mountain View}
  \country{USA}
}

\renewcommand{\shortauthors}{Chang et al.}

\begin{abstract}
Recommender systems often must maximize a primary objective while ensuring secondary ones satisfy minimum thresholds, or "guardrails." This is critical for maintaining a consistent user experience and platform ecosystem, but enforcing these guardrails despite orthogonal system changes is challenging and often requires manual hyperparameter tuning. We introduce the Automated Constraint Targeting (ACT) framework, which automatically finds the minimal set of hyperparameter changes needed to satisfy these guardrails. ACT uses an offline pairwise evaluation on unbiased data to find solutions and continuously retrains to adapt to system and user behavior changes. We empirically demonstrate its efficacy and describe its deployment in a large-scale production environment.
\end{abstract}

\maketitle

\section{Introduction}
Modern recommender systems optimize for multiple, often competing, objectives, such as user satisfaction, responsibility, and novelty ~\cite{zheng}. This multi-objective optimization is commonly recast as a constrained optimization problem: maximize a primary objective while ensuring secondary objectives satisfy specific constraints, or "guardrails" ~\cite{twostage}. In practice, enforcing these guardrails is crucial for managing the long-term consequences of recommendation decisions. From a system stability perspective, it ensures a consistent user experience and stable creator incentives (if applicable). From a developer velocity point of view, it is critical for determining the true impact of experimental changes, as shifting secondary objectives can obscure if primary objective gains are genuine or byproducts of such movements.

However, enforcing these guardrails is challenging. Seemingly unrelated system changes, such as adding features to a model or changing a model architecture, can cause metrics to drop below their required thresholds ~\cite{lrf}. While hyperparameters can be introduced to tune these objectives, finding the right values is time-consuming, must be done per-experiment, and scales poorly as the number of objectives grows.

To address this, we introduce the ~\textbf{Automated Constraint Targeting (ACT)} framework. ACT uses offline evaluations on randomized data to automatically find the minimal hyperparameter adjustments needed to ensure all secondary metric guardrails are met. We empirically show ACT's effectiveness and describe its deployment on YouTube.

\subsection{Related Work}


Much research has been done on choosing linear scalarization weights for multi-objective optimization~\cite{spotify,jeunen}; for example, Jeunen et al. \cite{jeunen} cast scalarization as a decision-making task that chooses weights to optimize for a North Star metric. Other research reframes multi-objective optimization as a constrained optimization task. For instance, Cai et al. ~\cite{twostage} propose regularizing a primary policy to balance secondary objectives.

We complement this research by focusing on satisfying guardrails on secondary objectives in the face of orthogonal system changes (e.g. model architecture changes) and in a per-experiment manner. This helps ensure that all experiments satisfy guardrails, thereby facilitating fair comparisons against prod and reducing the need for tuning. Another key difference is that ACT continuously adapts to system, corpus, and user behavior changes through dynamic offline evaluation; notably, if a secondary metric's guardrail is already satisfied by the original ranking, it naturally sets the corresponding hyperparameter to zero, avoiding unnecessary adjustment to the ranking score.

\section{Problem Statement}
\label{sec:problem}
We examine the problem of ranking a set of candidate videos. Given the following:
\begin{itemize}
\item Primary metric $P$
\item $n$ secondary metrics $S_1, ..., S_n$
\item Guardrail secondary metric levels $\epsilon_1, ..., \epsilon_n$
\item A hyperparameter vector $W = (w_1, ..., w_n)$ where $w_i$ is a tuning hyperparameter for secondary metric $S_i$
\item Candidate items to rank: $item_1, ..., item_k$
\item A formula (parameterized by $W$) used to generate a ranking score $r$ for each item: $f(item_1, ..., item_k; W) \to (r_1, ..., r_k)$
\item Given the above, let $S_i(W)$ refer to the secondary metric value achieved when ranking using hyperparameter vector $W$.
\end{itemize}

We can cast the problem as finding the smallest possible weights that satisfy the secondary metric guardrails:
\begin{gather*}
\min_{W} \|W\|_2^2 \\
\text{subject to} \quad S_i(W) \ge \epsilon_i \quad \text{for } i \in \{1, \dots, n\}
\end{gather*}
Minimizing the L2 norm of $W$ ensures we find the solution closest to the original scores ($W=0$) that satisfies the constraints.

\section{Methodology}
\subsection{Overview}\label{sec:methodology_overview}
\begin{figure*}[t]
\centering
  \centering
    \includegraphics[trim={0cm 0.1cm 0cm 0.1cm}, clip=true, width=0.8 \linewidth]{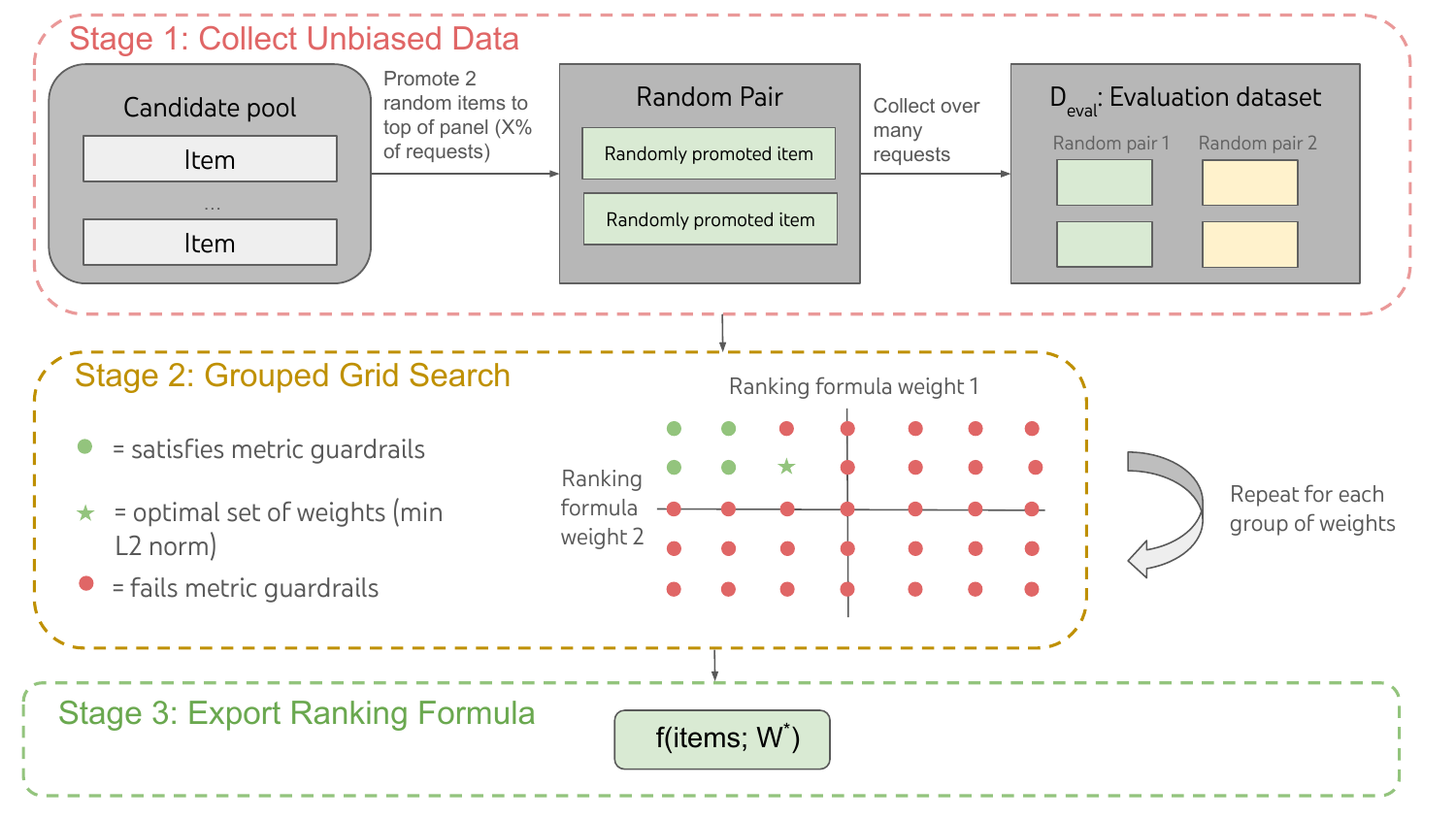}
    \caption{The ACT framework: 1) Random pairs of items are collected. 2) Weights are selected to satisfy guardrails. 3) The final weights $W^*$ are exported.}
    \label{fig:act}
\end{figure*}

The ACT framework searches for the optimal weights using a variant of grid search, guided by an offline evaluation metric that is predictive of online secondary metric movements.  Solving the constrained optimization problem in Section \ref{sec:problem} requires searching the n-dimensional space of $W$. The most robust approach is a multidimensional search that jointly optimizes all weights. To manage the computational cost, which grows exponentially with the number of objectives, ACT employs a \textbf{grouped grid search}. We partition the objectives into small, related groups (e.g., $\{S_1, S_2\}$, $\{S_3, S_4\}$). The algorithm then performs a grid search over the weights for each group jointly, while holding weights for other groups constant. This `block coordinate descent' approach correctly handles interactions between correlated objectives within a group, as illustrated in Figure~\ref{fig:act}.

\subsection{Data Preparation}
We generate random pairs data $D_{random} = \{(item_A^{(j)}, item_B^{(j)})\}_{j=1}^{m}$ consisting of $m$ pairs of items, by picking two random candidates from the nominated candidate pool and showing them at prominent positions in the recommendations panel for a small fraction of requests ~\cite{fairness}. Using randomized data is crucial for accurate off-policy evaluation, as it avoids the selection biases inherent in logged data from a production ranking policy~\cite{uw}. This data is then annotated with model predictions and ranking formula terms (for the formula described in Section~\ref{sec:problem}) specific to each experimental treatment.

\subsection{Pairwise Evaluation from Randomized Pair Data}
To find the optimal weights, we need an offline estimator for the true online metric $S_i(W)$, which we denote as $\hat{S}_i(W)$. To compute this estimator, we make the assumption that each metric $S_i$ has an item-level analogue $s_i$; for example, if $S_i$ represents the number of likes on videos over all video impressions, $s_i$ is whether an individual video is liked.

For a pair of items $(A, B)$ from $D_{random}$, the one with the higher score $r$ is the "winner." The offline metric $\hat{S}_i(W)$ is then computed as the mean value of the item-level attribute $s_i$ computed over all winners: $ \hat{S}_i(W) = \mathbb{E}_{(A,B) \sim D_{random}} [s_{i,A}*I_{r_A>r_B} + s_{i,B}*I_{r_B \ge r_A}]$,
where $I$ is the indicator function. This allows us to efficiently simulate the metric impact of any candidate set of weights $W$.

The pairwise setup can be viewed as offline evaluation for the  two-arm bandit randomly sampled from all items to be ranked : only one item is shown from each pair, and user feedback is observed on that choice. Compared to full-slate permutation approaches (e.g., \cite{li2011unbiased}), which can significantly degrade user experience, pairwise evaluation offers a practical trade-off between unbiasedness and usability.

\subsection{Weight Selection Algorithm}
With the pairwise evaluation defined, we can find the optimal weights. Our general grouped search approach is detailed in Algorithm \ref{alg:act}. For each group of objectives, we perform a grid search over their corresponding weights. From the points on the grid that satisfy all guardrails for that group (as measured by our offline estimator $\hat{S}_i$), we select the one with the minimum L2 norm.

\begin{algorithm}[t]
\caption{ACT Grouped Weight Selection}\label{alg:act}
\begin{algorithmic}[1]
\State \textbf{Input:} A partition of objectives into groups $G_1, ..., G_k$
\State \textbf{Initialize:} $W \gets (0, ..., 0)$
\For{each group $G_j$ in the partition}
    \State Let $W_j$ be the weights corresponding to objectives in $G_j$.
    \State Define a grid of candidate values for $W_j$ as an $\epsilon$-net on $R^n$.
    \State valid\_weights $\gets \emptyset$
    \ForAll{candidate $W_{j,cand}$ in the grid}
        \State $W_{temp} \gets $ current $W$ with $W_j$ set to $W_{j,cand}$
        \State is\_valid $\gets$ true
        \ForAll{metric $S_i$ in $G_j$}
            \If{$\hat{S}_i(W_{temp}) < \epsilon_i$}
                \State is\_valid $\gets$ false; \textbf{break}
            \EndIf
        \EndFor
        \If{is\_valid}
            \State add $W_{j,cand}$ to valid\_weights
        \EndIf
    \EndFor
    \State \Comment{Find valid point with minimum L2 norm}
    \State $W^*_j \gets \arg\min_{W' \in \text{valid\_weights}} \|W'\|_2^2$
    \State Update $W$ with the weights from $W^*_j$
\EndFor
\State \textbf{Output:} $W^* \gets W$
\end{algorithmic}
\end{algorithm}

\subsection{Export Ranking Formula}
Given the final weights $W^*$, the framework exports the final ranking formula $f(item_1,...,item_k; W^*)$ for serving. A separate, guardrailed formula is generated for the production treatment and for each experimental treatment, ensuring that all variants satisfy the same metric constraints. The whole process of data preparation, offline evaluation, and exporting the ranking formula is done on a recurring basis (e.g. daily). This allows the weights to adapt in response to system changes (e.g. ranking changes) and organic user behavior changes.


\section{Empirical Results}
ACT has been successfully launched for multiple secondary objectives on YouTube. Below, we answer key questions about its performance.

\subsection{How effectively does ACT enforce guardrails?}
Our experiments centered on a model predicting long-term user satisfaction. We tested ACT's ability to correct for drops in two secondary objectives: $S_1$, a video property, and $S_2$, which governs ranking of different formats. For both objectives, we established two underlying model versions:
\begin{itemize}
    \item Prod: The baseline production model.
    \item Decrease variant: An experimental version of Prod, modified in a way that is known to significantly decrease the secondary metric.
\end{itemize}
Each of these model versions was then configured using two distinct approaches for setting the relevant hyperparameter:
\begin{itemize}
    \item Fixed weight: The weight was set to a fixed human-defined value.
    \item ACT: The ACT framework was used to select the weight.
\end{itemize}
We then evaluated how well each method kept the metric from dropping in live experiments. Table~\ref{tb:experiments_combined} shows results from these experiments. Note that the metric changes for $S_1$ are an order of magnitude larger than for $S_2$; this is expected, as $S_1$ has higher natural variance. For reference, a routine retraining of the production model without other changes typically moves this metric by ~0.8\%. In both cases, ACT successfully moves the metric back towards neutral, while the fixed weight approach leaves the large metric shifts unaddressed.

\begin{table}[h]
\centering
\caption{ACT effectively counteracts metric drops. Values show relative change and 95\% confidence intervals.}
\label{tb:experiments_combined}
\begin{tabular}{l|cc|cc}
\toprule
\textbf{Model} & \multicolumn{2}{c|}{\textbf{Change in $S_1$}} & \multicolumn{2}{c}{\textbf{Change in $S_2$}} \\
\textbf{Variant} & \textbf{Fixed Wt.} & \textbf{ACT} & \textbf{Fixed Wt.} & \textbf{ACT} \\
\midrule
\multirow{2}{*}{Decrease} & -13.40\% & \textbf{-2.25\%} & +0.39\% & \textbf{+0.03\%} \\
 & \scriptsize{[-13.44\%, -13.35\%]} & \scriptsize{[-2.32\%, -2.18\%]} & \scriptsize{[+0.34\%, +0.44\%]} & \scriptsize{[-0.06\%, +0.11\%]} \\
\bottomrule
\end{tabular}
\end{table}

\subsection{Justifying the Sequential Simplification in Production}
\label{sec:validation}
While the general ACT framework supports jointly optimizing groups of correlated objectives, our production deployment uses a simplified sequential search where each objective is its own group. This is a practical choice driven by two factors: computational efficiency and the observed low correlation between our production guardrail metrics. Our path independence test provides strong evidence for this choice: when reversing the tuning order for $S_1$ and $S_2$, the weight chosen for $S_1$ was identical. In general, orthogonal secondary objectives do not need to be grouped together in the grouped grid search algorithm used by ACT.

\subsection{How predictive is the offline evaluation?}
Finally, the success of ACT hinges on its offline pairwise evaluation being predictive of online metric movements. Across 100 past A/B tests, we found a strong Pearson correlation of 0.82 between our offline estimator, $\hat{S}_1(W)$, and the observed online change in the true metric, $S_1(W)$, confirming that the core mechanism of ACT is reliable.

\begin{figure}[h]  
\centering
    \centering
    \includegraphics[width=0.7 \linewidth]{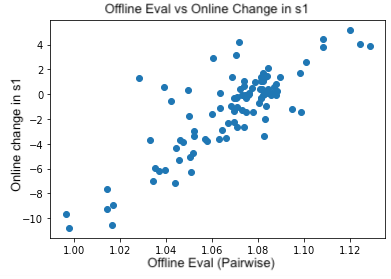}
    \caption{Relationship between pairwise offline evaluation and online $S_1$ movement, with Pearson correlation 0.82.}
    \label{fig:pairwise_offline}
\end{figure}

\section{Conclusion and Future Work}
We presented ACT, a framework for ensuring secondary metric guardrails in recommender systems by finding minimal hyperparameter adjustments. We reformulated the problem as a constrained optimization and introduced a general grouped search algorithm. We also described details of ACT's deployment in production.

Future work could explore more sample-efficient methods such as Bayesian optimization as an alternative to grid search. Another avenue is to develop heuristics for automatically grouping objectives based on their correlation.

\bibliographystyle{ACM-Reference-Format}
\bibliography{sigproc}

\end{document}